\documentclass[twocolumn,showpacs,prl,amsfonts,amsmath,amssymb,floatfix]{revtex4} 
\usepackage{color}
\usepackage{hhline}
\usepackage{mathrsfs}
\usepackage{graphicx}
\usepackage{dcolumn}
\usepackage{bm}
\usepackage{multirow,bigdelim}
\usepackage{booktabs}
\usepackage{ulem}
\input{epsf}
\usepackage{amsmath}

\def\Hscf1{{\cal H}_{SCF}}
\def\dvscf1{\Delta V_{SCF}}
\def\Hscfkq1{{\cal H}_{SCF}^{{\mathbf k}+{\mathbf q}} }
\def\kq1{{\mathbf k} + {\mathbf q}}
\def\k1{{\mathbf k}}
\def\kp1{{\mathbf k}'}
\def\q1{{\mathbf q}}
\def\rp1{{\mathbf r}'}
\def\thetan1{\tilde{\theta}_{F,n}}
\def\thetam1{\tilde{\theta}_{F,m}}
\def\thetamn1{\tilde{\theta}_{m,n}}
\def\thetanm1{\tilde{\theta}_{n,m}}
\def\g1{{\mathbf G}}
\def\gp1{{\mathbf G}'}

\begin{document}

\title{Effect of electron-phonon interactions on orbital fluctuations in iron-based superconductors}

\author{Yusuke Nomura$^{1}$}
\author{Kazuma Nakamura$^{2}$}
\author{Ryotaro Arita$^{1,3}$}
\affiliation{$^1$Department of Applied Physics, University of Tokyo, 7-3-1 Hongo, Bunkyo-ku, Tokyo, 113-8656, Japan} 
\affiliation{$^2$Quantum Physics Section, Kyushu Institute of Technology, 1-1 Sensui-cho, Tobata, Kitakyushu, Fukuoka, 804-8550, Japan}
\affiliation{$^3$JST-PRESTO, Kawaguchi, Saitama, 332-0012, Japan}
\date{\today}

\begin{abstract}
To investigate the possibility whether electron-phonon coupling can enhance orbital fluctuations in iron-based superconductors, 
we develop an {\it ab initio} method to construct the effective low-energy models including the phonon-related terms.
With the derived effective electron-phonon interactions and phonon frequencies, we estimate the static part ($\omega=0$) of the phonon-mediated effective on-site intra- or inter-orbital electron-electron attractions as $\sim-0.4$ eV and exchange or pair-hopping terms as $\sim-0.02$ eV. 
We analyze the model with the derived interactions together with the Coulomb repulsions within the random phase approximation. We find that 
the enhancement of the orbital fluctuations due to the electron-phonon interactions is small, and that 
the spin fluctuations enhanced by the Coulomb repulsions dominate.
It leads to the superconducting state with the sign reversal in the gap functions ($s_\pm$-wave). 
\end{abstract} 
\maketitle 

{\it -Introduction}.
The mechanism of superconductivity in iron-based superconductors has attracted much attention owing to its high critical temperature ($T_c$)~\cite{Fe-review}. The pairing symmetry of the Cooper pair is a central issue and in active debate.
There are two strong candidates. One is the spin-fluctuation-mediated $s_{\pm}$-pairing with a sign reversal in the gap functions~\cite{Fe-Mazin, Fe-Kuroki,Fe-Ikeda,Fe-Chubukov,Fe-Kemper,Fe-Fernandes,Fe-third,Fe-FRG,Fe-Yao,Fe-Arita}, 
which is consistent with the phase sensitive experiments~\cite{Hanaguri,Chen}. 
The other is the orbital-fluctuation-mediated $s_{++}$-pairing without sign changes~\cite{Fe-Kontani1,Fe-Kontani2,Fe-niigata}, which seems to be compatible with the robustness of the superconductivity against impurity doping~\cite{imp_exp,Fe-imp}.  
As for the orbital fluctuations, it has recently been proposed that not only the Coulomb interactions but also the electron-phonon (el-ph) couplings can play a role~\cite{Fe-Kontani1,Fe-Kontani2}.
To examine the scenario quantitatively and conclude the controversy on the pairing symmetry, 
it is highly required to derive, from first principles, the effective model both with the electronic and the phononic part and analyze it. While the {\it ab initio} derivations of the electronic model have widely been done~
\cite{cRPA-ex6, cRPA-ex7,Fe-UAG,Fe-2D,Fe-Hirayama,misawa}, that for the phonon-related part has not been performed due to the lack of methodology.


In this Letter, we present an {\it ab initio} effective low-energy model including phonon terms for the iron-based superconductor, LaFeAsO. The effective el-ph interactions and phonon frequencies in the model are estimated using the density-functional perturbation theory (DFPT)~\cite{DFPT} with a constraint that screening processes in the Fe-$3d$ bands are excluded. From the derived parameters, 
we estimated the phonon-mediated on-site electron-electron (el-el) attractions. The resulting values for the static part are $\sim -0.4$ eV for the intra- and inter-orbital terms and $\sim -0.02$ eV for the exchange and pair-hopping ones. 
The magnitude of the obtained exchange interaction is considerably smaller than the one which gives the $s_{++}$-wave solution $\sim -0.4$ eV~\cite{Fe-Kontani1}. 

We analyzed the model including electronic repulsions as well as the derived phonon-mediated interactions within the random phase approximation (RPA). Due to the small phonon-mediated on-site exchange and pair-hopping interactions, the enhancement of the orbital fluctuations is small, and the spin fluctuations enhanced by Coulomb repulsions are dominant and mediate the $s_{\pm}$-wave pairing.  

{\it -Method}.
Here, we describe the {\it ab initio} downfolding method to evaluate the el-ph couplings and the phonon frequencies in the effective model. 
The model consists of the phonons and the electronic degrees of freedom belonging to the subspace near the Fermi level, which we call target subspace ($t$-subspace). In the case of LaFeAsO, we choose the Hilbert space spanned by the Fe-$3d$ bands as the $t$-subspace.  
In this low-energy model, the degrees of freedom residing far from the Fermi level are eliminated, which give the renormalization for the effective parameters~\cite{Kotliar_review,Imada_review}. 
We consider the renormalization effects by calculating partially-screened (renormalized) el-ph couplings and phonon frequencies with excluding the $t$-subspace contribution, which is to be accounted when the model is solved~\cite{ferdi,Bauer,BKBO}.
This exclusion is achieved by imposing a constraint to the DFPT calculation. 
Below, we describe the basic idea and practical treatments.

The frequency for phonon mode $\nu$ with momentum ${\bf q}$ is determined by the secular equation 
$ \sum_{\kappa' \alpha' }  ( C_{\kappa \kappa'} ^ {\alpha \alpha'}({\bf q})  
 -   M_{\kappa}  \omega_{{\bf q }\nu} ^2   \delta_{\kappa \kappa'} \delta_{\alpha \alpha'} )   e _{\kappa'}^{\alpha'}  ({\bf q \nu})  = 0 $ with $M_{\kappa}$ and $\alpha$ being the mass of atom $\kappa$ and the Cartesian components, respectively.
The interatomic force constants $C_{\kappa \kappa'} ^ {\alpha \alpha'}({\bf q})$ are given by $C_{\kappa \kappa'} ^ {\alpha \alpha'}({\bf q}) = \phantom{}^{\rm{bare}}C_{\kappa \kappa'} ^ {\alpha \alpha'}({\bf q}) + \phantom{}^{\rm{ren.}}C_{\kappa \kappa'} ^ {\alpha \alpha'}({\bf q})$, where 
\begin{eqnarray}
\label{Eq:SE_ph}
\phantom{}^{\rm{ren.}} C_{\kappa \kappa'} ^ {\alpha \alpha'}({\bf q}) =  \frac{1}{N}
  \int \biggl(  \frac{\partial n (\bf r) }{ \partial u^{ \alpha}_{\kappa}  ({\bf q})} \biggr ) ^ {\ast} 
 \frac {\partial V_{\rm{ion}} (\bf r)} { \partial u^{  \alpha'}_{\kappa'} ({\bf q}) }  d{\bf r}
\end{eqnarray}
with $N$, $n({\bf r})$, $u({\bf q})$, and $V_{\rm{ion}}({\bf r})$ being the number of unit cells in the crystal, electron density, ionic displacement, and ionic potential, respectively.
Here, $\phantom{}^{\rm{ren.}}C_{\kappa \kappa'} ^ {\alpha \alpha'}({\bf q})$ gives the renormalization of the phonon frequencies via the linear el-ph coupling, and  $\phantom{}^{\rm{bare}}C_{\kappa \kappa'} ^ {\alpha \alpha'}({\bf q})$ gives the bare phonon frequencies~\cite{note_Cbare}. 
The el-ph couplings are evaluated as 
$g_{n^{\prime} n}^{\nu}({\bf k, q})  =  \sum_{\kappa\alpha}
   e^{\alpha}_{\kappa} ({\bf q}\nu )  d^{\kappa \alpha}_{n'n} ({\bf k}, {\bf q}) / \sqrt{2 M_{\kappa} \omega_{{\bf q} \nu}}$, where $d^{\kappa \alpha}_{n'n} ({\bf k}, {\bf q}) = 
   \left\langle  \psi_{n' \kq1} \left|  \partial V_{\rm{SCF}} ({\bf r}) / \partial u^{\alpha}_{\kappa} ({\bf q})   \right| \psi_{n\k1} \right\rangle$ is a coupling between the Bloch states $\psi_{n{\bf k}}$ with momentum {\bf k} and band $n$ and $\psi_{n'{\bf k+q}}$.
The derivative of the self-consistent field potential $\partial V_{\rm{SCF}} ({\bf r}) / \partial  u^{\alpha}_{\kappa}({\bf q}) $ 
is written as 
\begin{eqnarray} 
\label{Eq:delV}
 \frac{\partial V_{\rm{SCF}} ({\bf r})}{\partial u^{\alpha}_{\kappa}({\bf q}) } 
  =   \frac{\partial V_{\rm{ion}} ({\bf r}) }{\partial  u^{\alpha}_{\kappa} ({\bf q}) }  + 
\int  \biggl( \frac { e^2 } {|{\bf r}-{\bf r^{\prime}} | }   \nonumber \\ 
   + \frac{dV_{\rm{xc}}({\bf r})}{dn}  \delta ({\bf r - \bf r'})     \biggr )
    \frac{\partial n ({\bf r'}) }{\partial u^{\alpha}_{\kappa}({\bf q}) }
 d {{\bf r}'}
\end{eqnarray}
with $V_{\rm{xc}}({\bf r})$ being the exchange-correlation potential.
In the r.h.s. of this formula, the first term denotes the bare potential and the second one denotes the screening potential. 
The electron density response $\partial n ({\bf r}) / \partial  u^{\alpha}_{\kappa}({\bf q})$ in Eqs. (\ref{Eq:SE_ph}) and (\ref{Eq:delV}) gives the renormalization of the phonon frequencies and the screening for 
the el-ph couplings.
This response is explicitly written as
 \begin{eqnarray}
 \label{Eq:dens-res}
\! \! \! \! \frac{\partial n ({\bf r}) }{\partial u^{\alpha}_{\kappa}({\bf q}) } \!\!=\!2\!\!\sum_{nm{\bf k}}  \!  \frac{f_{n {\bf k}} \! \!   - \! \! f_{m {\bf k+q} } } { \epsilon_{n {\bf k}}  \! \! - \!  \!  \epsilon_{m {\bf k+q}} }
   \psi_{n \bf k}^{\ast} ({\bf r}) \psi_{m {\bf k+q}} ({\bf r})
  d^{\kappa \alpha}_{mn}({\bf k}, {\bf q})\!, \end{eqnarray}
where $\epsilon_{n {\bf k}}$ and $f_{n {\bf k}}$ are the eigenvalue and its occupancy, respectively.

For the derivation of the effective model, we calculate the density response with excluding the contribution from the case where both $\psi_{m {\bf k+q}}$ and $\psi_{n {\bf k}}$ belong to the $t$-subspace.  Then, with the resulting density response, we evaluate the partially-screened (renormalized) quantities such as $g^{(p)}$ and $\omega^{(p)}$. We call the scheme ``constrained DFPT". 
Without the constraint, fully-screened quantities are calculated, to which we attach the superscript ``$f$'', instead of ``$p$"~\cite{note_supple}. 
 
\begin{figure}[htbp]
\vspace{0cm}
\begin{center}
\includegraphics[width=0.42\textwidth]{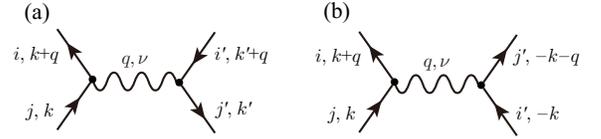}
\caption{Feynman diagrams for phonon-mediated effective (a) el-el [Eq.~(\ref{V})] and (b) pairing [Eq.~(\ref{Vp})] interactions. Solid lines with arrows are electron propagators, wavy lines are phonon Green's functions, and dots represent el-ph couplings.} 
\label{fig_diagram}
\end{center}
\end{figure}

Now we write down the phonon-related terms in the effective model. The effective el-ph interactions are 
\begin{eqnarray}
\! \! \! \! {\cal H}_{\rm el\mathchar`-ph} \! =  \!  \frac{1}{\sqrt{N_{\k1}}} 
\sum_{\q1\nu}  \! \sum_{\k1ij\sigma} 
g^{(p) \nu}_{i j}(\k1,\q1) c_{i\kq1}^{\sigma \dagger } c_{j\k1}^{\sigma} 
( b_{\q1\nu}\!+\!b^{\dagger}_{-\q1\nu}\!),
\end{eqnarray}
where $b_{\q1\nu}$ ($b^{\dagger}_{\q1\nu}$) is the annihilation (creation) operator of the phonon with the wave vector $\q1$ and the branch $\nu$. $c_{i\k1}^{\sigma}$ ($c_{i\k1}^{\sigma\dagger}$) annihilates (creates) the $i$-th Wannier orbital's electron with the wave vector $\k1$ and the spin $\sigma$. $N_{\k1}$ is the number of $\k1$ points. The phonon one-body part is given as
\begin{eqnarray}
{\cal H}_{\rm ph} = \sum_{\q1\nu}   \omega^{(p)}_{\q1\nu}  b^{\dagger}_{\q1 \nu} b_{\q1 \nu}.
\end{eqnarray}
The momentum-space-averaged phonon-mediated effective el-el interaction 
$V^{(p)}_{ij , i'j'}$ [Fig.~\ref{fig_diagram}(a)] is given by
\begin{eqnarray}
V^{(p)}_{ij , i'j'}(\omega_l) =   \frac{1}{N_{\q1}}  \sum_{\q1\nu}  
&& \left( \frac{1}{N_{\k1}} \sum_{\k1}  g^{(p) \nu}_{i j }(\k1,\q1)  \right ) 
  D_{\q1 \nu} ^{(p)}(\omega_l) \nonumber  \\
&  \times &
  \left [  \frac{1}{N_{\k1}}  \sum_{\kp1} \left( g^{(p)\nu}_{i' j'}(\kp1, \q1) \right )^{\ast}  \right ],  \label{V}
\end{eqnarray} 
where $\omega_l = 2  \pi l T$ is the boson Matsubara frequency and $D_{\q1 \nu} ^{(p)}(\omega_l) = - 2  \omega^{(p)}_{\q1\nu} / ( \omega_l^2 + \omega^{(p)2}_{\q1\nu} ) $.
Note that $V^{(p)}_{ij , i'j'}$ corresponds to the on-site quantity because of the momentum-space averaging.
This $V^{(p)}_{ij , i'j'}$ is distinguished from the momentum-space-averaged phonon-mediated effective pairing interaction $V'^{(p)}_{ij , i'j'}$ [Fig.~\ref{fig_diagram}(b)] as 
\begin{eqnarray}
V'^{(p)}_{ij , i'j'}(\omega_l) =   \frac{1}{N_{\q1}N_{\k1}}  \sum_{\q1\nu}  \sum_{\k1}
&& \left(   g^{(p) \nu}_{i j }(\k1,\q1)  \right ) 
  D_{\q1 \nu} ^{(p)}(\omega_l) \nonumber \\ 
  &  \times &
   \left( g^{(p)\nu}_ {j' i'}(\k1, \q1) \right )^{\ast}. \label{Vp}
\end{eqnarray} 

\begin{table*}
\caption[t]{
Our calculated static phonon-mediated effective electron-electron interaction $V_{ij,i'j'}(\omega_{l}=0)$ and pairing interaction $V'_{ij,i'j'}(\omega_{l}=0)$. Note that the values are represented with the negative sign. The upper (lower) panel shows the partially (fully) screened interactions.  $V_{ij,i'j'}$ is symmetric with respect to $i \leftrightarrow j$, $i' \leftrightarrow j'$, and $(ij) \leftrightarrow (i'j')$. $V'_{ij,i'j'}$ is symmetric with respect to $(ii') \leftrightarrow (jj')$ and $(ij) \leftrightarrow (i'j')$. Units are given in eV.}
\begin{center}
\begin{tabular}{c ccccc c cccc c@{\   } ccccc  c cccc c cccc }
\hline\hline 
&\multicolumn{5}{c} {$-V^{(p)}_{ii,jj}$} & &\multicolumn{4}{c}{$-V^{(p)}_{ij,ij} (= -V^{(p)}_{ij,ji}) \times 10$ }&
&\multicolumn{5}{c} {$-V'^{(p)}_{ii,jj}$} & &\multicolumn{4}{c}{$-V'^{(p)}_{ij,ji}$} 
& &\multicolumn{4}{c}{$-V'^{(p)}_{ij,ij} \times 10 $}\\
\cline{2-6}\cline{8-11}\cline{13-17}\cline{19-22}\cline{24-27}
&1& 2&3&4&5  &&2&3&4&5   &&1& 2&3&4&5  &&2&3&4&5  &&2&3&4&5\\
\hline 
1& 0.47  &  0.44 & 0.44 & 0.38 & 0.44   &&  \ 0.23 \   & \ 0.23 \  & \ 0.21 \ & \ 0.03 \  
&& 0.57 &  0.49 & 0.49 & 0.33 & 0.44    &&  0.09  & 0.09 & 0.22 & 0.12     && 0.001  & 0.001 &  \ 0.41 & \ 0.61 \\
2&  -       &  0.43 & 0.41 & 0.36 & 0.42    &&    -       & 0.06 & 0.20 & 0.23
&&  -       &  0.52 & 0.48 & 0.27 & 0.46    &&    -       & 0.10 & 0.16 & 0.14     &&    -     & -0.21 & -0.21& \ 0.28  \\
3&  -       &     -     & 0.43 & 0.36 & 0.42    &&    -        &    -    & 0.20 & 0.23 
&&  -       &     -     & 0.52 & 0.27 & 0.46    &&    -        &    -    & 0.16 & 0.14     &&    -     &      -    & -0.21&\  0.28    \\ 
4&  -       &     -     &   -      & 0.32 & 0.35    &&    -        &   -     &   -      & 0.03 
&&  -       &     -     &   -     & 0.54  & 0.23   &&    -         &   -     &   -      & 0.08     &&    -    &      -    &   -       &-0.03 \\
5&  -       &     -     &   -      &    -     & 0.43   &&     -        &   -     &   -      &   -    
&&  -      &     -      &   -     &    -     & 0.69   &&    -         &   -     &   -     &   -         &&    -      &  -       &   -       &    -    \\ 
\hline \hline \\
\end{tabular}

\begin{tabular}{c ccccc c cccc c ccccc  c cccc c cccc }
\hline\hline 
&\multicolumn{5}{c} {$-V^{(f)}_{ii,jj} \times10$} & &\multicolumn{4}{c}{$-V^{(f)}_{ij,ij} (= -V^{(f)}_{ij,ji}) \times  10$ }&
&\multicolumn{5}{c} {$-V'^{(f)}_{ii,jj}\times 10$} & &\multicolumn{4}{c}{$-V'^{(f)}_{ij,ji}$} 
& &\multicolumn{4}{c}{$-V'^{(f)}_{ij,ij} \times 10 $}\\
\cline{2-6}\cline{8-11}\cline{13-17}\cline{19-22}\cline{24-27}
&1& 2&3&4&5  &&2&3&4&5   &&1& 2&3&4&5  &&2&3&4&5  &&2&3&4&5\\
\hline 
1& 0.27  &  0.28 & 0.28 & 0.24 & 0.21   &&  \ 0.25 \   & \ 0.25 \  & \ 0.24 \ & \ 0.01 \  
&& 1.80 &  0.97 & 0.97 & -0.46 &\  0.32&&  0.13  & 0.13 & 0.31 & 0.17     && -0.11  &-0.11&  \ 0.86 & \ 0.98  \\
2&  -       &  0.37 & 0.24 & 0.27 & 0.21    &&    -       & 0.03 & 0.14 & 0.18 
&&  -       &  1.60 & 1.05 & -0.89 & \ 1.07  &&    -       & 0.15 & 0.22 & 0.18     &&    -     & -0.30 & -0.57& \ 0.14  \\
3&  -       &     -     & 0.37 & 0.27 & 0.21    &&    -        &    -    & 0.14 & 0.18 
&&  -       &     -     & 1.60 & -0.89 & \ 1.07  &&    -        &    -    & 0.22 & 0.18     &&    -     &      -    & -0.57&\  0.14    \\ 
4&  -       &     -     &   -      & 0.39 & 0.11    &&    -        &   -     &   -      & 0.04 
&&  -       &     -     &   -     &\ 3.48  & -1.87  &&    -         &   -     &   -      & 0.12     &&    -    &      -    &   -       & -0.02 \\
5&  -       &     -     &   -      &    -     & 0.28   &&     -        &   -     &   -      &   -    
&&  -      &     -      &   -     &    -     &\ 4.33   &&    -         &   -     &   -     &   -         &&    -      &  -       &   -       &    -    \\ 
\hline \hline 
\end{tabular}
\end{center}
\label{tab_elph}
\end{table*}

{\it -Results}.
We performed density-functional calculations with {\sc quantum espresso} package~\cite{QE}. The generalized-gradient approximation (GGA) with the Perdew-Burke-Ernzerhof parameterization~\cite{PBE} and the Troullier-Martins norm-conserving pseudopotentials~\cite{TM} in the Kleinman-Bylander representation~\cite{KB} are adopted.
The cutoff energy for the wave functions is set to 95 Ry, 
and we employ 8$\times$8$\times$6 ${\mathbf k}$ points. The phonon frequencies and the el-ph interactions are calculated using the DFPT~\cite{DFPT} with and without the constraint, where 4$\times$4$\times$3 ${\mathbf q}$-mesh and a Gaussian smearing of 0.02 Ry are employed. The maximally localized Wannier function~\cite{maxloc} is used as the basis of the model. The lattice parameter and the internal coordinates are fully optimized and we get $a$ = 4.0344 ${\rm \AA}$, $c$ = 8.9005 ${\rm \AA}$, $z_{\rm La}$ = 0.14233, and $z_{\rm As}$ = 0.63330. These values are in good agreement with those of Refs. \cite{Fe-Boeri,Fe-Singh_ph}.

\begin{figure}[htbp]
\vspace{0cm}
\begin{center}
\includegraphics[width=0.48\textwidth]{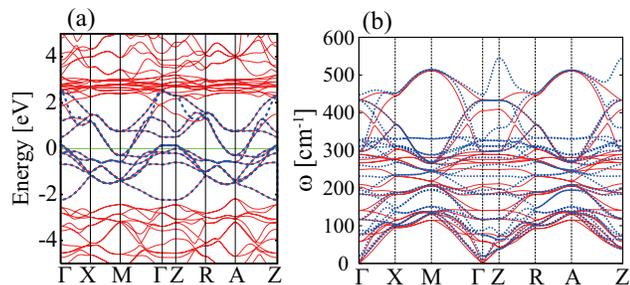}
\caption{(Color online) (a) Band structure of LaFeAsO for the optimized structure (red solid curves). Blue dotted curves denote the Wannier-interpolated band dispersion. (b) Fully (red solid curves) and partially (blued dotted curves) renormalized phonon dispersion of LaFeAsO.}
\label{fig_band}
\end{center}
\end{figure} 

We show in Fig.~\ref{fig_band}(a) our calculated GGA band (solid curves) of LaFeAsO with the optimized structure, and compare with the Wannier-interpolated band (dotted ones) for the Fe-$3d$ orbitals. Hereafter, $d_{3Z^2-R^2}$, $d_{XZ}$, $d_{YZ}$, $d_{X^2-Y^2}$, and $d_{XY}$ orbitals are represented as 1, 2, 3, 4, and 5, respectively, where the $X$ and $Y$ axes are parallel to the nearest Fe-As bonds and the $Z$ axis is perpendicular to the FeAs layer. 
The screening and self-energy effects within the energy range from the bottom of the Fe-$3d$ bands up to 2.32 eV are excluded to derive $g^{(p)}$ and $\omega^{(p)}$.


Figure~\ref{fig_band}(b) displays our calculated phonon dispersions with (dotted curves) and without (solid ones) the constraint on the $t$-subspace screening. 
We see a discernible difference in the frequencies for the phonon modes which couple to the $t$-subspace electrons. 
However, the difference is not large, at most $\sim$ 20 percent.


Table~\ref{tab_elph} lists our calculated static on-site phonon-mediated interactions. The upper left-side two 5$\times$5 matrices are the el-el interaction $V^{(p)}_{ij,i'j'}(0)$ in Eq.~(\ref{V}). 
The intra- and inter-orbital terms $V^{(p)}_{ii,jj}(0)$ are $\sim$$-$0.4 eV, while the 
exchange and pair-hopping terms $V^{(p)}_{ij,ij}(0)=V^{(p)}_{ij,ji}(0)$ are rather small as $\sim$$-$0.02 eV. Compared to the on-site Coulomb repulsion $U$$\sim$2 eV~\cite{misawa}, $V^{(p)}_{ii,ii}(0)$$\sim$$-$0.4 eV are not negligible. However, it should be noted here that, while $U(\omega_l)$ is 
almost constant
up to the typical plasmon frequency ($\sim$ 25 eV in the iron-based superconductors~\cite{Fe-dynamic}),
the attractions $V^{(p)}_{ij,i'j'}(\omega_l)$ quickly decay
as $\omega_l$ increases and vanish around $\omega_l \sim \omega_D$ with $\omega_D$ being the Debye frequency. 


The three matrices in the upper right side of Table~\ref{tab_elph} are the effective pairing interactions $V'^{(p)}_{ij,i'j'}(0)$ in Eq.~(\ref{Vp}). 
Due to the off-site pairing interactions, the pair-hopping terms $V'^{(p)}_{ij,ji}(0)$ are substantially larger in magnitude than the on-site quantities $V^{(p)}_{ij,ji}(0)$.

The lower part of Table~\ref{tab_elph} describes fully screened ones $V^{(f)}_{ij,i'j'}(0)$ and $V'^{(f)}_{ij,i'j'}(0)$.
The intra- and inter-orbital terms are efficiently screened from the $t$-subspace electrons, while others not.
We note that the quantity $\sum_{ij}V'^{(f)}_{ij,ji} N_i(0)N_j(0)/N(0)$ with $N_i(0)$ ($N(0)$) being the partial (total) density of states at the Fermi level is $\sim 0.18$, which gives a reasonable estimate to the total el-ph coupling constant $\lambda \sim 0.2$ in this system~\cite{Fe-Boeri}.  


{\it -Effect on pairing symmetry}.
Here we analyze a five-band model including the electronic repulsions and phonon-mediated interactions within the RPA. The calculation detail follows Refs.~\cite{Fe-Kontani1,Fe-Kontani2}. The spin and charge susceptibilities are given by
$\hat{\chi}^{s(c)}(q) = \hat{\chi}^{0}(q) [1- \hat{\Gamma}^{s(c)}\hat{\chi}^{0}(q) ]^{-1}$,
where $\hat{\chi}^{0}(q)$ is the irreducible susceptibility and $\Gamma^{s}_{ij,i'j'}= U$, $U'$, $J$, and $J'$ for $i$ = $j$ = $i'$ = $j'$,  $i$ = $i'$ $\neq$ $j$ = $j'$,   $i$ = $j$ $\neq$ $i'$ = $j'$,  $i$ = $j'$ $\neq$ $i'$ = $j$, respectively~\cite{Fe-Kontani1}. $U$ $(U')$ is the intra- (inter-) orbital Coulomb repulsion, $J$ is the Hund's coupling, and $J'$ is the pair-hopping.
The matrix $\hat{\Gamma}^{c}$ is given by $\hat{\Gamma}^{c}=-\hat{C}-2\hat{V}^{(p)}(\omega_l)$, where $C_{ij,i'j'}= U$, $-U'+2J$, $2U'-J$, and $J'$ for $i$ = $j$ = $i'$ = $j'$,  $i$ = $i'$ $\neq$ $j$ = $j'$,   $i$ = $j$ $\neq$ $i'$ = $j'$,  $i$ = $j'$ $\neq$ $i'$ = $j$, respectively~\cite{Fe-Kontani1}. 

With these susceptibilities, we solve the linearized gap equation
\begin{eqnarray}
\lambda_E \Delta_{ii'}(k) =&& \frac{T}{N}\sum_{k',j_i} W_{ij_1,j_4i'} (k-k') G^{0}_{j_1j_2}(k') \Delta_{j_2,j_3} (k') \nonumber \\
&&\times G^{0}_{j_4,j_3}(-k'),
\label{eq_gapeq}
\end{eqnarray}
where $\Delta_{ii'}(k)$ ($G^{0}_{ii'}(k)$) is the gap (non-interacting Green's) function in the orbital representation and $W_{ij,i'j'}(q)$ is the pairing interaction kernel. For the singlet pairing, $\hat{W}(q)$ = $-\frac{3}{2}\hat{\Gamma}^{s}  \hat{\chi}^{s}(q)  \hat{\Gamma}^{s}  + \frac{1}{2} \hat{\Gamma}^{c}   \hat{\chi}^{c}(q) \hat{\Gamma}^{c}  -\frac{1}{2} (\hat{\Gamma}^{s} - \hat{\Gamma}'^{c}) $ with $\hat{\Gamma}'^{c}=-\hat{C}-2\hat{V}'^{(p)}(\omega_l)$~\cite{triplet}. The eigenvalue $\lambda_E$ grows as the temperature decreases, reaching unity at the superconducting transition temperature. 

We adopt two dimensional model and 64 $\times$ 64 ${\mathbf k}$-point meshes and 2048 Matsubara frequencies are taken. The temperature and the filling are set to $T = 0.02$ eV and $n=6.1$, respectively.  We discuss the structure of the diagonal elements of the gap-function matrices in the band representation at the lowest Matsubara frequency and we denote them as $\phi_m(\mathbf k)$ with $m$ being the band index.
When the Coulomb interactions are large, the RPA treatment is known to be unstable~\cite{Fe-Kuroki}. So, we scale the original {\it ab initio} electronic interactions $U$, $U'$, $J$ and $J'$~\cite{note2} by 1/2 with keeping the phonon-mediated interactions $V^{(p)}_{ij,i'j'}$ and $V'^{(p)}_{ij,i'j'}$ at the original values in Table~\ref{tab_elph}. 

\begin{figure}[tbp]
\vspace{0cm}
\begin{center}
\includegraphics[width=0.48\textwidth]{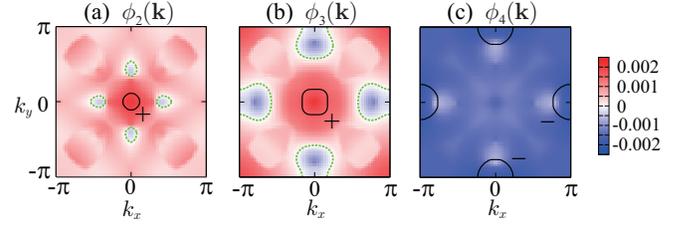}
\caption{(Color online) RPA results for the gap functions (a) $\phi_2$,  (b) $\phi_3$, and (c) $\phi_4$. We used $T=0.02$ eV and $n=6.1$. As for the interaction parameters, we scaled the {\it ab initio} Coulomb interactions in Ref.~\cite{misawa} by 1/2, while we used the original values in Table~\ref{tab_elph} for the phonon-mediated el-el interactions. The black-solid (green-dotted) curves represent the FS's (nodes of the gap functions), together with the sign of gap functions on each FS's.} 
\label{fig_gap}
\end{center}
\end{figure}

Our calculated gap functions $\phi_2(\k1)$, $\phi_3(\k1)$, and $\phi_4(\k1)$ are shown in Figs.~\ref{fig_gap}(a), (b), and (c), respectively. We see the sign change in the gap functions on the Fermi surfaces (FS's), i.e., the $s_{\pm}$-wave state is realized. 
In our parameter setting, the phonon-mediated interactions are considerably overemphasized to the scaled electronic repulsions; nevertheless, we get the $s_{\pm}$-wave solution. Thus, the $s_{++}$-wave pairing based on the orbital fluctuations due to the el-ph interactions would not be realized in the {\it ab initio} parameter range~\cite{note_param}.

To clarify why the $s_{\pm}$-wave state are stable even if we introduce the el-ph interactions, we analyze simpler models. We use the following Coulomb parameters: $U = 0.8$ eV, $U' = 0.69 U$, and $J = J' = 0.16 U$. For the phonon-mediated interaction matrix $\hat{V}$~\cite{note3}, we consider two parameter sets. One is the same as that of Ref.~\cite{Fe-Kontani1}; the exchange terms $V_{ij,ij}$ are set to be equal to the intra-orbital ones $V_{ii,ii}$ such as $V_{24,24} = V_{34,34} = V_{22,22} = V_{33,33} = -V_{22,33} = V(\omega_l)$, where $V(\omega_l) = V(0) \omega^2_D / (\omega^2_l + \omega^2_D)$ with $V(0) = -0.385$ eV and $\omega_D=0.02$ eV~\cite{note_234}. 
Note that $V_{ij,i'j'}$ has the symmetry on the index interchange as $i \leftrightarrow j$, $i' \leftrightarrow j'$, and $(ij) \leftrightarrow (i'j')$. 
The other parameter set is based on the present {\it ab initio} results; the exchange terms are appreciably weakened from the intra- and inter-orbital ones; $V_{24,24} = V_{34,34} = V(\omega_l) / 20$ and $V_{22,22} = V_{33,33} = V_{22,33} = V(\omega_l)$. 
With the former parameter set, we see 
an enhancement of the orbital fluctuations and get the $s_{++}$-wave state, while, in the latter case, the spin fluctuations enhanced by the Coulomb repulsions 
develop and bring about the $s_{\pm}$-wave pairing.
In the analysis for the former case, we checked that the sign change of $V_{22,33}$ has no qualitative effect on the pairing symmetry.
Therefore, it is deduced that the magnitudes of $V_{24,24}$ and $V_{34,34}$ are crucial  in determining $s_{++}$ or $s_{\pm}$-wave states.  

When the exchange and pair-hopping terms $V_{24,24}=V_{24,42} = V_{34,34} = V_{34,43}$ are large in magnitude, the scattering channels $W_{24,42}$ and $W_{34,43}$, which connect $\Delta_{44}$ and $\Delta_{22}$, and $\Delta_{44}$ and $\Delta_{33}$, respectively, are enhanced through the charge sector and both channels have the positive value~\cite{Fe-Kontani1}. These attractive pairing channels lead to the same sign in $\Delta_{22}$, $\Delta_{33}$, and $\Delta_{44}$. Since the FS's consist of these three (2-4) orbitals, the $s_{++}$-wave state is realized. 
On the other hand, when $V_{24,24}$ and $V_{34,34}$ are small as with the {\it ab initio} results, the scattering channels enhanced by the spin part become dominant and the $s_{\pm}$-wave state is realized. 
Thus, the el-ph interactions with the {\it ab initio} energy scale alone cannot drive the orbital-fluctuation-mediated $s_{++}$-wave pairing~\cite{note_VC}.

{\it -Conclusion}.
We have developed an {\it ab initio} downfolding method for el-ph coupled systems and applied it to the derivation of the effective model of LaFeAsO. With the derived effective el-ph interactions $g^{(p)}$ and phonon frequencies $\omega^{(p)}$, we have estimated the phonon-mediated effective on-site el-el interactions as $V^{(p)}_{ii,ii}(0)\sim V^{(p)}_{ii,jj}(0)\sim-0.4$ eV and $V^{(p)}_{ij,ij}(0)=V^{(p)}_{ij,ji}(0)\sim-0.02$ eV. 
We have analyzed the derived five band model consisting of Fe-$3d$ bands using the RPA. The $s_{++}$-wave pairing is not realized with the {\it ab initio} el-ph interaction with the tiny exchange and pair-hopping terms,
and the $s_{\pm}$-wave state mediated by spin-fluctuations is robustly realized.


While our study is concentrated on LaFeAsO, it would be interesting to perform a comprehensive cDFPT study for the other iron-based superconductors including those with different topology of Fermi surfaces~\cite{KFeSe1,KFeSe2,KFeSe3,KFeSe4},
and investigate the material dependence.  Furthermore, 
our developed cDFPT also enables the quantitative study of other superconductors and different areas of research related to phonons, such as multiferroics, thermoelectric materials, dielectrics, and polaron problems.
These applications are interesting and important future issues.

\begin{acknowledgements}
{\it -Acknowledgments}.
We would like to thank Hiroshi Kontani, Seiichiro Onari, Hiroaki Ikeda, Ryosuke Akashi and Hideyuki Miyahara for fruitful discussions. 
This work was supported by Funding Program for World-Leading Innovative R\&D on Science and Technology
(FIRST program) on "Quantum Science on Strong Correlation" JST-PRESTO, Grants-in-Aid for Scientific Research (22740215, 22104010, 23110708, 23340095, 23510120, 25800200) and the Next Generation Super Computing Project and Nanoscience Program from MEXT,
Japan.
Y. N. is supported by Grant-in-Aid for JSPS Fellows (12J08652).
\end{acknowledgements}

\clearpage 

{\LARGE Supplemental  Material}

\renewcommand{\theequation}{S.\arabic{equation}}
\setcounter{equation}{0}
\renewcommand{\tablename}{Table S}

\noindent
\section{ S.1 The form of $^{\rm{bare}}C^{\alpha\alpha'}_{\kappa\kappa'}({\bf q})$}
In the density functional perturbation theory (DFPT), the interatomic force constant $C_{\kappa \kappa'} ^ {\alpha \alpha'}({\bf q})$ is given by~\cite{DFPT} 
\begin{eqnarray}
\label{Eq:C}
C_{\kappa \kappa'} ^ {\alpha \alpha'}({\bf q})  =  \frac{1}{N}
 \biggl [ &&
\frac{\partial^2 E_{\rm{N}} }
{\partial u^{\ast\alpha}_{\kappa} ({\bf q})\partial u^{\alpha'}_{\kappa'} ({\bf q})}  \nonumber \\
&+& \int n({\bf r})\frac{\partial^2 V_{\rm{ion}}({\bf r})} 
{\partial u^{\ast\alpha}_{\kappa} ({\bf q})\partial u^{\alpha'}_{\kappa'} ({\bf q})} d{\bf r}  \nonumber \\ 
&+&  \int \biggl(  \frac{\partial n (\bf r) }{ \partial u^{ \alpha}_{\kappa}  ({\bf q})} \biggr ) ^ {\ast}  
 \frac {\partial V_{\rm{ion}} (\bf r)} { \partial u^{  \alpha'}_{\kappa'} ({\bf q}) }  d{\bf r} \biggr ]
\end{eqnarray}
where $E_{\rm{N}}$ is the energy for the Coulomb interactions between different nuclei: 
\begin{eqnarray}
E_{\rm{N}} = \frac{e^2}{2} \sum_{\kappa,\kappa'} \frac{Z_{\kappa}Z_{\kappa'}}{| {\bf R}_{\kappa}  - {\bf R}_{\kappa'}  |}
\end{eqnarray}
with $Z_{\kappa}$ and ${\bf R}_{\kappa}$ being the charge and the position of $\kappa$-th nucleus, respectively. 
In the r.h.s. of Eq. (\ref{Eq:C}), the first term describes the ionic contribution and the second (third) term describes the contribution from the quadratic (linear) electron-phonon (el-ph) coupling. 
Since the present effective model includes the linear el-ph couplings that can renormalize the phonon frequencies after the model analysis,
we define (ionic contribution) + (contribution from the quadratic coupling) as ``bare'' term, and (contribution from the linear coupling) as ``renormalizing" term.
The explicit form of $^{\rm{bare}}C^{\alpha\alpha'}_{\kappa\kappa'}({\bf q})$ is
\begin{eqnarray}
\label{Eq:Cbare}
\phantom{}^{\rm{bare}}C^{\alpha\alpha'}_{\kappa\kappa'}({\bf q})  =  \frac{1}{N}  \biggl [ &&
\frac{\partial^2 E_{\rm{N}}}
{\partial u^{\ast\alpha}_{\kappa} ({\bf q})\partial u^{\alpha'}_{\kappa'} ({\bf q})}
 \nonumber \\ 
&+&   \int    n({\bf r})\frac{\partial^2 V_{\rm{ion}}({\bf r})} 
{\partial u^{\ast\alpha}_{\kappa} ({\bf q})\partial u^{\alpha'}_{\kappa'} ({\bf q})} d{\bf r} \biggr ].
\end{eqnarray}

\noindent
\section{ S.2 Comparison between ${\rm \bf c}$DFPT and ${\rm \bf c}$RPA}

Here, we 
compare
the present constrained DFPT with
the constrained random phase approximation (cRPA)~\cite{ferdi}. 
The cRPA is the standard method for deriving effective Coulomb parameters in the model.

In the constrained method, in general, one divides the one-particle Hilbert space into two parts. One is the target subspace ($t$-subspace) consisting of electronic bands near the Fermi level, which are the target-band degrees of freedom of the effective model. The other is the rest of the Hilbert space, which is referred to as $r$-subspace. 
Within the RPA,
the total irreducible polarization $\chi^0$ can be divided into $\chi^0_t$ and $\chi^0_r$ with $\chi^0_t$ being the polarization within the $t$-subspace and $\chi^0_r = \chi^0 - \chi^0_t$ is the rest of the polarization. 

In the cRPA to derive the effective electron-electron interactions in the model, the screening of the Coulomb interaction is decomposed into the two steps~\cite{ferdi}; 
\begin{eqnarray}
W^{(p)}=\left(1-v\chi^0_r\right)^{-1}v
\end{eqnarray}
and 
\begin{eqnarray}
W^{(f)}=\left(1-W^{(p)}\chi^0_t\right)^{-1}W^{(p)},  
\end{eqnarray}
where $v$ is the bare Coulomb interaction. Such a decomposition holds even though $v$ is replaced by $\tilde{v}=v+K_{\rm{xc}}$ with $K_{\rm{xc}}=\delta V_{\rm{xc}}/\delta n$ defined as the exchange-correlation kernel. Here, $V_{\rm{xc}}$ and $n$ are the exchange-correlation potential and the electron density, respectively~\cite{note1}. Then, we obtain 
\begin{eqnarray}
\tilde{W}^{(p)}=\left(1-\tilde{v}\chi^0_r\right)^{-1}\tilde{v} \label{Wp}
\end{eqnarray}
and 
\begin{eqnarray}
\tilde{W}^{(f)}=\left(1-\tilde{W}^{(p)}\chi^0_t \right)^{-1}\tilde{W}^{(p)}. \label{Wf} 
\end{eqnarray}

Now, the cDFPT to derive the phonon-related term in the effective model is formulated as follows: First, on the basis of the usual DFPT scheme~\cite{DFPT}, the induced electron density $\Delta n$ to the perturbation $\Delta V_{\rm{ion}}$ (bare potential) is given by 
\begin{eqnarray} 
\label{Eq:deln}
\Delta n &=&  \underbrace{ \chi^0  \left(1-\tilde{v}\chi^0\right)^{-1}}_{\chi_{\rm{LDA}}}  \Delta V_{\rm{ion}}  \\  
\label{Eq:s2-KN}
&=& \chi^0 \Delta V_{\rm{SCF}},  
\end{eqnarray} 
where the change in the self-consistent field potential $\Delta V_{\rm{SCF}}$ (screened potential) is written as 
\begin{eqnarray}
\label{Eq:s3-KN}
\Delta V_{\rm{SCF}}=\left(1-\tilde{v}\chi^0 \right)^{-1}\Delta V_{\rm{ion}}. 
\end{eqnarray}
Note Eqs. (\ref{Eq:s2-KN}) and (\ref{Eq:s3-KN}) correspond to Eqs. (3) and (2) in the main text, respectively.
Since the el-ph coupling $g$ is given as the matrix element of the electron scattering via $\Delta V_{\rm{SCF}}$,
the same decomposition as Eqs.~(\ref{Wp}) and (\ref{Wf}) holds for the fully screened el-ph interaction; that is, $g^{(f)}=\left(1-\tilde{v}\chi^0\right)^{-1}g^{(b)}$ is decomposed as 
\begin{eqnarray}
\label{Eq:gp}
g^{(p)}=\left(1-\tilde{v}\chi^0_r\right)^{-1}g^{(b)}       
\end{eqnarray}
and
\begin{eqnarray}
\label{Eq:gf}
g^{(f)}=\left(1-\tilde{W}^{(p)}\chi^0_t\right)^{-1}g^{(p)}.        
\end{eqnarray}
Therefore, the present cDFPT is formally based on the cRPA-like decomposition, but the difference is that, in the former, $\tilde{v}$ is used instead of $v$.

The similar idea is applied to the derivation of the phonon frequencies in the effective model. In this case, the self-energy is decomposed. 
One can show that Eq. (1) in the main text is rewritten as
\begin{eqnarray}
^{\rm{ren.}}C = | g'^{(b)} | ^2 \chi_{\rm{LDA}}, 
\end{eqnarray}
where $g'^{(b)} = \sqrt{2M \omega^{(b)}}g^{(b)}$ with the bare phonon frequency $\omega^{(b)}$.
In this expression, we omit the subscripts for simplicity. 
Then, we define the phonon self-energy in the DFPT scheme as 
\begin{eqnarray}
\label{Eq.sigma}
\Sigma = \frac{\phantom{}^{\rm{ren.}}C}{2M \omega^{(b)} }= | g^{(b)} | ^2 \chi_{\rm{LDA}}
\end{eqnarray}
This self-energy can be divided into two contributions as $\Sigma = \Sigma_t + \Sigma_r$. 
Here, $\Sigma_r = | g^{(b)} | ^2 \chi^{r}_{\rm{LDA}}$ with $\chi^{r}_{\rm{LDA}} = \chi^0_{r}  \left(1-\tilde{v}\chi^0_{r} \right)^{-1}$  is the phonon self-energy due to the interactions between the $r$-subspace electrons and the phonons. 
The interactions between the $t$-subspace electrons and the phonons through the partially-screened coupling $g^{(p)}$ give rise to the self-energy $\Sigma_t = | g^{(p)} | ^2 \chi^{t}_{\rm{LDA}}$ with $\chi^{t}_{\rm{LDA}} = \chi^0_{t}  \left(1-\tilde{W}^{(p)}\chi^0_{t} \right)^{-1}$.  
The decomposition of $\Sigma$ into $\Sigma_t$ and $\Sigma_r$ is achieved by dividing the band sum in Eq.~(3) in the main text into the target-target contribution and the others.
Then, the partially-dressed phonon Green's function $D^{(p)}$ is given by
\begin{eqnarray}
 [ D^{(p)} ]^{-1} = [D^{(b)}] ^{-1} - \Sigma_{r}, 
\end{eqnarray}
where $D^{(b)}$ is the bare phonon Green's function and its pole position gives the bare phonon frequency $\omega^{(b)}$. Similarly, the pole of $D^{(p)}$ gives the effective phonon frequency $\omega^{(p)}$ in the model. If we further consider $\Sigma_t$, the fully-dressed phonon Green's function $D^{(f)}$ is derived as 
\begin{eqnarray}
 [ D^{(f)} ]^{-1} = [D^{(p)}] ^{-1} - \Sigma_{t}. 
\end{eqnarray}


\begin{thebibliography}{999}
\bibitem{Fe-review} G. R. Stewart, Rev. Mod. Phys. {\bf 83}, 1589 (2011).

\bibitem{Fe-Mazin}I. I. Mazin, D. J. Singh, M. D. Johannes, and M. H. Du, Phys. Rev. Lett. {\bf 101}, 057003 (2008).
\bibitem{Fe-Kuroki}K. Kuroki, S. Onari, R. Arita, H. Usui, Y. Tanaka, H. Kontani, and H. Aoki, Phys. Rev. Lett. {\bf 101}, 087004 (2008). 
\bibitem{Fe-Ikeda}H. Ikeda, J. Phys. Soc. Jpn. {\bf 77}, 123707 (2008).
\bibitem{Fe-Chubukov}A. V. Chubukov, D. V. Efremov, and I. Eremin, Phys. Rev. B {\bf 78}, 134512 (2008).
\bibitem{Fe-third}T. Nomura, J. Phys. Soc. Jpn. {\bf 78}, 034716 (2009).
\bibitem{Fe-FRG}F. Wang, H. Zhai, Y. Ran, A. Vishwanath, and D.-H. Lee, Phys. Rev. Lett. {\bf 102}, 047005 (2009).
\bibitem{Fe-Yao}Z.-J. Yao, J.-X. Li, and Z. D. Wang, New J. Phys. {\bf 11}, 025009 (2009).
\bibitem{Fe-Arita}R. Arita and H. Ikeda, J. Phys. Soc. Jpn. {\bf 78}, 113707 (2009).
\bibitem{Fe-Kemper}A. F. Kemper, T. A. Maier, S. Graser, H.-P. Cheng, P. J. Hirschfeld, and D. J. Scalapino, New J. Phys. {\bf 12}, 073030 (2010).
\bibitem{Fe-Fernandes}R. M. Fernandes and J. Schmalian,  Phys. Rev. B {\bf 82}, 014521 (2010).

\bibitem{Hanaguri}T. Hanaguri, S. Niitaka, K. Kuroki, H. Takagi, Science {\bf 328}, 474 (2010).
\bibitem{Chen}C.-T. Chen, C. C. Tsuei, M. B. Ketchen, Z.-A. Ren, Z. X. Zhao, Nature Physics {\bf 6}, 260 (2010).

\bibitem{Fe-Kontani1}H. Kontani and S. Onari, Phys. Rev. Lett. {\bf 104}, 157001 (2010). 
\bibitem{Fe-Kontani2}T. Saito, S. Onari, and H. Kontani, Phys. Rev. B {\bf 82}, 144510 (2010).
\bibitem{Fe-niigata}Y. Yanagi and Y. Yamakawa, Y. \=Ono,  Phys. Rev. B {\bf 81}, 054518 (2010).
\bibitem{imp_exp}Y. Kobayashi, A. Kawabata, S.-C. Lee, T. Moyoshi, and M. Sato, J. Phys. Soc. Jpn. {\bf 78}, 073704 (2009); M. Sato, Y. Kobayashi, S.-C. Lee, H. Takahashi, E. Satomi, and Y. Miura, {\it ibid}. {\bf 79}, 014710 (2010).
\bibitem{Fe-imp}S. Onari and H. Kontani, Phys. Rev. Lett. {\bf 103}, 177001 (2009).

\bibitem{cRPA-ex6}K. Nakamura, R. Arita, and M. Imada, J. Phys. Soc. Jpn. {\bf 77}, 093711 (2008).
\bibitem{Fe-UAG}T. Miyake, L. Pourovskii, V. Vildosola, S. Biermann, and A. Georges, J. Phys. Soc. Jpn. Suppl. C {\bf 77}, 99 (2008).
\bibitem{cRPA-ex7}T. Miyake, K. Nakamura, R. Arita, and M.Imada, J. Phys. Soc. Jpn. {\bf 79}, 044705 (2010). 

\bibitem{Fe-2D}K. Nakamura, Y. Yoshimoto, Y. Nohara, and M. Imada, J. Phys. Soc. Jpn. {\bf 79} 123708 (2010).
\bibitem{misawa} T. Misawa, K. Nakamura, and M. Imada, Phys. Rev. Lett. {\bf 108}, 177007 (2012).

\bibitem{Fe-Hirayama}M. Hirayama, T. Miyake, and M. Imada, Phys. Rev. B {\bf 87}, 195144 (2013).

\bibitem{DFPT}S. Baroni, S. de Gironcoli, A. Dal Corso, and P. Giannozzi, Rev. Mod. Phys. {\bf 73}, 515 (2001).
\bibitem{Kotliar_review}G. Kotliar, S. Y. Savrasov, K. Haule, V. S. Oudovenko, O. Parcollet, and C. A. Marianetti, Rev. Mod. Phys. {\bf 78}, 865 (2006).
\bibitem{Imada_review}M. Imada and T. Miyake, J. Phys. Soc. Jpn. {\bf 79}, 112001 (2010).


\bibitem{ferdi}F. Aryasetiawan, M. Imada, A. Georges, G. Kotliar, S. Biermann, and A. I. Lichtenstein, Phys. Rev. B {\bf 70}, 195104 (2004).
\bibitem{Bauer}J. Bauer, J. E. Han, and O. Gunnarsson, Phys. Rev. B {\bf 84}, 184531 (2011).
\bibitem{BKBO}R. Nourafkan, F. Marsiglio, and G. Kotliar, Phys. Rev. Lett. {\bf 109}, 017001 (2012).
\bibitem{note_Cbare} The explicit form of $\phantom{}^{\rm{bare}}C_{\kappa \kappa'} ^ {\alpha \alpha'}({\bf q})$ is given in Supplemental Material (Ref.~\cite{supple}).
\bibitem{supple} See Supplemental Material at http://@@@.
\bibitem{note_supple}The effective parameters for Coulomb interactions are widely derived using the constrained random phase approximation (cRPA)~\cite{ferdi}. The comparison between the present cDFPT and the cRPA is given in Supplemental Material in Ref.~\cite{supple}. 

\bibitem{QE} P. Giannozzi, S. Baroni, N. Bonini, M. Calandra, R. Car,
C. Cavazzoni, D. Ceresoli, G. L. Chiarotti, M. Cococcioni,
I. Dabo, A. Dal Corso, S. de Gironcoli, S. Fabris, G. Fratesi,
R. Gebauer, U. Gerstmann, C. Gougoussis, A. Kokalj,
M. Lazzeri, L. Martin-Samos, N. Marzari, F. Mauri, R. Mazzarello,
S. Paolini, A. Pasquarello, L. Paulatto, C. Sbraccia, S. Scandolo,
G. Sclauzero, A. P. Seitsonen, A. Smogunov, P. Umari, and
R. M. Wentzcovitch, J. Phys.: Condens. Matter {\bf 21}, 395502 (2009);
http://www.quantum-espresso.org/.

\bibitem{PBE}
J. P. Perdew, K. Burke, and M. Ernzerhof, Phys. Rev. Lett. {\bf 77}, 3865 (1996).
\bibitem{TM}
N. Troullier and J. L. Martins, Phys. Rev. B {\bf 43}, 1993 (1991). 
\bibitem{KB}
L. Kleinman and D. M. Bylander, Phys. Rev. Lett. {\bf 48}, 1425 (1982). 
\bibitem{maxloc}N. Marzari and D. Vanderbilt, Phys. Rev. B {\bf 56}, 12847 (1997); 
I. Souza, N. Marzari, and D. Vanderbilt, {\it ibid}. {\bf 65}, 035109 (2001).

\bibitem{Fe-Boeri}L. Boeri, O. V. Dolgov, and A. A. Golubov, Phys. Rev. Lett. {\bf 101}, 026403 (2008).
\bibitem{Fe-Singh_ph}
D. J. Singh and M.-H. Du, Phys. Rev. Lett. {\bf 100}, 237003 (2008).

\bibitem{Fe-dynamic}
P. Werner, M. Casula, T. Miyake, F. Aryasetiawan, A. J. Millis, and S. Biermann, Nature Physics {\bf 8}, 331 (2012).

\bibitem{triplet} In Refs.~\cite{chiral_p_1,chiral_p_2}, the possibility of the triplet pairing was also pursued.
\bibitem{chiral_p_1}A. Aperis, P. Kotetes, G. Varelogiannis, and P. M. Oppeneer, Phys. Rev. B {\bf 83}, 092505 (2011). 
\bibitem{chiral_p_2}A. Aperis and G. Varelogiannis, arXiv:1303.2231.
\bibitem{note2} As for the effective Coulomb interactions, we employ the values in Ref.~\cite{misawa}.

\bibitem{note_param}We note that the $s_{\pm}$-wave state is always realized when we change the scaling parameter for Coulomb interactions from 0.4 to 0.6.


\bibitem{note3} We use the common phonon-mediated interaction parameters for $\hat{\Gamma}^{c}$ and $\hat{\Gamma}'^{c}$, i.e., $\hat{\Gamma}^{c} = \hat{\Gamma}'^{c}=-\hat{C}-2\hat{V}(\omega_l)$. 
\bibitem{note_234}For simplicity, the $V$ terms are introduced only for the 2, 3, and 4 orbitals which compose the FS's, following Ref.~\cite{Fe-Kontani1}. 
\bibitem{note_VC}We find that el-ph interactions do not enhance the orbital fluctuations, but 
it was recently proposed that the vertex corrections could possibly 
enhance them~\cite{Fe-SCVC,Fe-Miyahara}.
\bibitem{Fe-SCVC} S. Onari and H. Kontani, Phys. Rev. Lett. {\bf 109}, 137001 (2012). 
\bibitem{Fe-Miyahara} H. Miyahara, R. Arita, H. Ikeda, Phys. Rev. B {\bf 87}, 045113 (2013). 
\bibitem{KFeSe1}J. Guo, S. Jin, G. Wang, S. Wang, K. Zhu, T. Zhou, M. He, and X. Chen, Phys. Rev. B {\bf 82}, 180520(R) (2010).
\bibitem{KFeSe2}Y. Zhang, L. X. Yang, M. Xu, Z. R. Ye, F. Chen, C. He, H. C. Xu, J. Jiang, B. P. Xie,	 J. J. Ying, X. F. Wang, X. H. Chen, J. P. Hu, M. Matsunami, S. Kimura, and D. L. Feng, Nature Materials {\bf 10}, 273 (2011). 
\bibitem{KFeSe3}T. Qian, X.-P. Wang, W.-C. Jin, P. Zhang, P. Richard, G. Xu, X. Dai, Z. Fang, J.-G. Guo, X.-L. Chen, and H. Ding, Phys. Rev. Lett. {\bf 106}, 187001 (2011).
\bibitem{KFeSe4}L. Zhao, D. Mou, S. Liu, X. Jia, J. He, Y. Peng, L. Yu, X. Liu, G. Liu, S. He, X. Dong, J. Zhang, J. B. He, D. M. Wang, G. F. Chen, J. G. Guo, X. L. Chen, X. Wang, Q. Peng, Z. Wang, S. Zhang, F. Yang, Z. Xu, C. Chen, and X. J. Zhou, Phys. Rev. B,  {\bf 83}, 140508(R) (2011).

\end{thebibliography}

\begin{thebibliography}{99}
\bibitem[S1]{DFPT}S. Baroni, S. de Gironcoli, A. Dal Corso, and P. Giannozzi, Rev. Mod. Phys. {\bf 73}, 515 (2001).


\bibitem[S2]{ferdi}
F. Aryasetiawan, M. Imada, A. Georges, G. Kotliar, S. Biermann, and A. I. Lichtenstein, Phys. Rev. B {\bf 70}, 195104 (2004).


\bibitem[S3]{note1}Strictly speaking, this expression [Eq. (\ref{Eq:deln})] is valid only when the ionic potential $V_{\rm{ion}}$ is local. In practice, we utilize the pseudopotential, which has non-local part. In this case, we have to introduce three-point response functions, however, it does not change the outline presented in this section.

\end{thebibliography}
\end{document}